\documentclass{article}

\usepackage{arxiv}

\usepackage[utf8]{inputenc} 
\usepackage[T1]{fontenc}    
\usepackage{hyperref}       
\usepackage{url}            
\usepackage{booktabs}       
\usepackage{amsfonts}       
\usepackage{nicefrac}       
\usepackage{microtype}      
\usepackage{graphicx}
\usepackage{natbib}
\usepackage{doi}
\usepackage{amsmath}
\usepackage{array} 
\usepackage{multirow}
\usepackage{float}

\usepackage{caption}
\usepackage{threeparttable}

\title{Efficacy of Temporal Fusion Transformers for Runoff Simulation}

\date{} 

\author{Sinan Rasiya Koya\thanks{corresponding author} \\
	Civil and Environmental Engineering\\
	University of Nebraska-Lincoln\\
	\texttt{ssinanrk2@huskers.unl.edu} \\
	\And
	Tirthankar Roy \\
	Civil and Environmental Engineering\\
	University of Nebraska-Lincoln\\
	\texttt{roy@unl.edu} \\
}

\renewcommand{\shorttitle}{TFT for Runoff Simulation}

\hypersetup{
    pdftitle={Efficacy of Temporal Fusion Transformers for Runoff Simulation},pdfauthor={Sinan Rasiya Koya, Tirthankar Roy},
    pdfkeywords={Temporal Fusion Transformers, LSTM, Machine Learning, Caravan Dataset, CAMELS Dataset}
    }

\begin{document}
\maketitle

\begin{abstract}
Combining attention with recurrence has shown to be valuable in sequence modeling, including hydrological predictions. Here, we explore the strength of Temporal Fusion Transformers (TFTs) over Long Short-Term Memory (LSTM) networks in rainfall-runoff modeling. We train ten randomly initialized models, TFT and LSTM, for 531 CAMELS catchments in the US. We repeat the experiment with five subsets of the Caravan dataset, each representing catchments in the US, Australia, Brazil, Great Britain, and Chile. Then, the performance of the models, their variability regarding the catchment attributes, and the difference according to the datasets are assessed. Our findings show that TFT slightly outperforms LSTM, especially in simulating the midsection and peak of hydrographs. Furthermore, we show the ability of TFT to handle longer sequences and why it can be a better candidate for higher or larger catchments. Being an explainable AI technique, TFT identifies the key dynamic and static variables, providing valuable scientific insights. However, both TFT and LSTM exhibit a considerable drop in performance with the Caravan dataset, indicating possible data quality issues. Overall, the study highlights the potential of TFT in improving hydrological modeling and understanding. 
\end{abstract}


\section{Introduction}
Hydrology has been witnessing a paradigm shift since the introduction of machine learning into the field. Various models can be trained to learn the underlying dynamics in a hydrological system. It is argued that, given a large amount of data, machines can learn processes that are unknown to humans \citep{Nearing2021}. In the past decade, several machine learning algorithms have been proven successful in hydrological modeling and prediction \citep{Kratzert2018, RasiyaKoya2024a}. Most commonly, the family of sequence prediction algorithms aligns with the type of problems hydrological modelers deal with. In the initial research, simple fully connected artificial neural networks (ANN) have been applied. However, it was the recurrent neural networks, specifically the long short-term memory (LSTM) networks, that outperformed the existing benchmarks of conceptual and physics-based hydrological models. Later, with the introduction of attention-based models, which transformed the natural language processing field, several hydrologists studied their performance in hydrological modeling tasks \citep{Castangia2023, Liu2024a, Yin2022a}.

In our recent study \citet{RasiyaKoya2024a}, we have found that combining recurrence with attention would significantly improve streamflow prediction (or forecast) rather than applying them independently. We trained three models for each in 2,610 catchments worldwide and compared their performance. Three models were 1) LSTM, a recurrent neural network, 2) Transformers, an attention-based neural network, 3) Temporal Fusion Transformers (TFT), a neural network architecture that combines aspects of recurrence and attention. Recurrent-based models and attention-based models have their own advantages over each other. While recurrent models consist of the Markovian nature (or state-evolution over time), similar to the hydrological processes, attention-based models can access the entire sequence at a time and learn from it without depending on just the previous step. LSTMs outperform Transformers in the mentioned work, primarily due to the dominant effect of state evolution over time element, a conclusion confirmed by another recent study \citet{Liu2024a}. At the same time, TFT outperforms both LSTM and transformers in streamflow prediction, as it has the advantages of both recurrence and attention. We also demonstrated the ability of TFTs to learn from distant data and predict the features of the hydrograph, such as rising and falling limbs. Additionally, the explainability of TFTs makes them more favorable.

Although TFTs have proven preferable for streamflow forecasting, their ability in rainfall-runoff modeling is yet to be established. Rainfall-runoff modeling, the simulation of the transformation of rainfall into runoff, is an equally essential problem to explore in hydrology besides streamflow forecasting. While streamflow prediction is concerned with the future states of flow in the rivers, rainfall-runoff modeling focuses on understanding and simulating the processes leading to the flow in the river. Therefore, rainfall-runoff modeling is desired to study long-term changes and gain insights into the fundamental hydrological behaviors of a region of interest.

In this study, we investigate the ability of TFTs in rainfall-runoff modeling. We compare the performance with the existing benchmark of LSTM trained with CAMELS. We assess the performance of models with respect to catchment attributes. The same experiments are repeated on five subsets of the Caravan dataset, and the performance is analyzed. Subsequently, we demonstrate the ability of TFT to learn from long sequences. Lastly, we show the explainability of the model and how we can gain insights into the rainfall-runoff processes and the quality of datasets. 

\section{Methods}
\subsection{Data and study region}
We use two catchment aggregated datasets in this study. Firstly, the Catchment Attributes and Meteorological data set for Large Sample Studies (CAMELS) data \citep{Addor2017, Newman2015} for training the TFT consistent with the existing benchmark experiment (see section 2.3). CAMELS dataset, primarily developed to support large-sample hydrological studies, consists of catchment attributes of 671 catchments across the United States and hydrometeorological time series aggregated for each catchment. The catchment attributes include topographic, climatic, hydrological, soil, vegetation, and geological characteristics. The hydrometeorological time series, spanning from 1st January 1980 to 31st December 2008, is derived from NLDAS \citep{Xia2012a}, Daymet \citep{Thornton2014}, and \citet{Maurer2002}. The dataset also provides the USGS daily streamflow at the outlet of each catchment. A summary of CAMELS data variables is given in Table B1 (Appendix B).

Secondly, we use the Caravan dataset \citep{Kratzert2023a}, derived by standardizing multiple catchment scale datasets worldwide, which include CAMELS datasets for different regions (US, Great Britain, Brazil, Australia, and Chile), LArge-SaMple DAta for Hydrology and Environmental Sciences for Central Europe (LamaH-CE), and Hydrometeorological Sandbox - École de technologie supérieure (HYSETS) \citep{Alvarez-Garreton2018, Arsenault2020, Chagas2020, Coxon2020, Fowler2021, Klingler2021}. Similar to CAMELS, the Caravan dataset also provides catchment attributes, hydrometeorological time series, and recorded outlet streamflow data for 6,830 catchments globally. However, in Caravan, the hydrometeorological variables are derived from ERA5-Land \citep{Munoz-Sabater2021}, and the catchment attributes are expanded with HydroATLAS \citep{Linke2019, Lehner2019}. A summary of Caravan data variables used in this study is given in Table C1C2 (Appendix C). 

\subsection{Models}
Two models are implemented in this study: LSTM and TFT. LSTMs are a variety of recurrent neural networks \citep{Hochreiter1997}. The key components in LSTM architecture are memory cells and gates. While the input gates regulate the amount of new information added to the memory cell, the forget gate controls the removal. Based on the memory content, the output gates determine the information to be output. The information in a sequence is processed step-by-step, in a Markovian manner, with state evolution over time inside the LSTMs. This property is a key merit of LSTMs in hydrological modeling, as the hydrological processes occur in a similar manner.

TFT is a specific model architecture (Figure 1) developed by combining the recurrent aspect of LSTMs and the attention feature of transformers \citep{Lim2021}. Here, the fundamental components are an LSTM block and, over that, an interpretable multihead self-attention block, which is known for its effectiveness in capturing long-range dependencies sequential data. In addition, there are gating mechanisms to learn better the complex patterns in the dataset, variable selection networks to scale the important variables at each time step, and static covariate encoders to assimilate static features. A more detailed description of TFT, in light of its application in hydrology, can be found in \citet{RasiyaKoya2024a}.

\begin{figure}[h]
	\centering
    \includegraphics[width=0.7\textwidth]{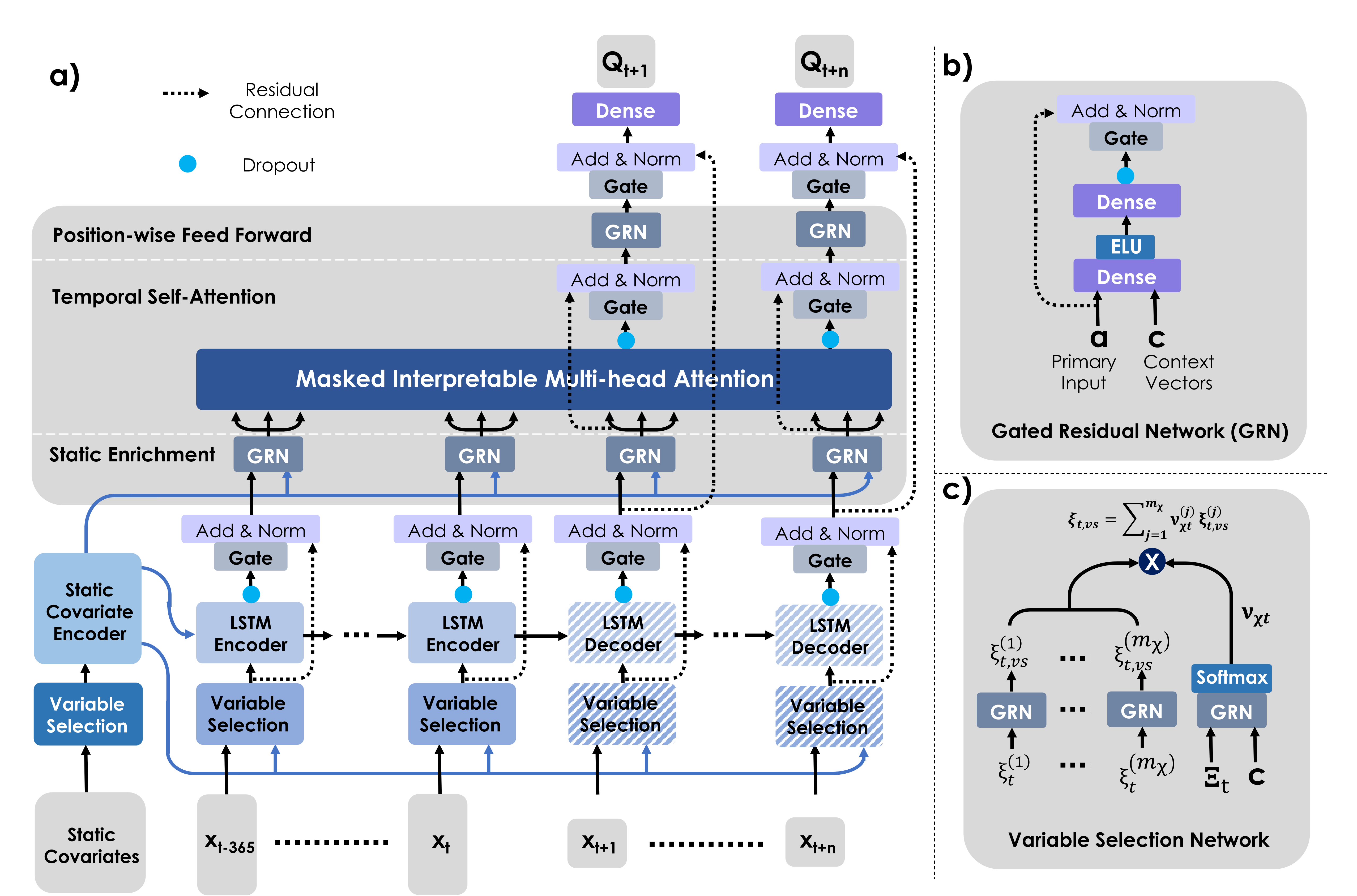}
    \caption{a) TFT architecture. Two key components of TFT are b) Gated Residual Networks (GRN) for effective processing of information and c) variable selection networks for filtering features from input (source: \citet{RasiyaKoya2023a}, adapted from \citet{Lim2021}}
	\label{fig:fig1}
\end{figure} 

\subsection{Benchmarks}

Most previous rainfall runoff ML model benchmarking studies have been done on CAMELS catchments. Among these studies, the best-reported performance is by \citep{Kratzert2021}. They trained ten LSTM, each initialized with different random seeds, and ensembled the outputs using equal-weight averaging. In their experiments, following \cite{Newman2017}, models were trained with all possible combinations of three meteorological datasets, provided simultaneously, for each catchment. Their results show that using multiple meteorological datasets (the NLDAS, Maurer, and Daymet time series, three meteorological datasets provided by CAMELS) improves the performance compared to using just one. Thus, providing all three time series subsets in the CAMELS achieved the best results for LSTM. The performance was compared to the Sacramento Soil Moisture Accounting (SAC-SMA) conceptual hydrological model, calibrated using dynamically dimension search \citep[DDS]{Tolson2007} with the same dataset. SAC-SMA achieved a median (across 531 CAMELS catchments) NSE of 0.705, whereas the LSTM achieved 0.821, outperforming the former.

\subsection{Experiments}
We trained TFTs following the exact training configuration to enable a fair comparison with the performance bar set by previous studies and benchmark TFTs. The experiment involves training a single model for all catchments, where the trained parameter set is shared across all catchments. Models are trained using the CAMELS dataset (of 531 catchments) from 1st October 1999 to 30th September 2008 and tested using that from 1st October 1989 to 30th September 1999 (referred to as the CAMELS experiment). The period before that (from 1st October 1980 to 30th September 1989) is used for validation. All models are trained for 30 epochs using Adam optimizer with a learning rate of 0.001 for the first ten, 0.0005 for the second ten, 0.0001 for the last ten epochs. We use the catchment-averaged NSE as the loss function \citep{Kratzert2019}. We did a grid search on TFT hyperparameters to find the combination that can give the optimum performance. In this process, we trained TFTs with combinations given in Table A1 (Appendix A) and selected the best combination based on the highest validation KGE. Subsequently, the simulated streamflow is compared with the observed to report the evaluation metrics (Table 1).

However, the CAMELS dataset might not be representative of all catchments across the globe since it only consists of catchments that belong to the contiguous US. This region only consists of specific hydrologic regimes, which include only a certain pattern of precipitation, streamflow, and catchment characteristics peculiar to the region. These regimes can be different in different regions worldwide. Additionally, regarding geographic diversity and climatic variability, the contiguous US and, thus, the CAMELS dataset represents only a limited spectrum. Therefore, we trained the LSTMs and TFTs on each subset of the Caravan dataset (referred to as the Caravan experiment), representing a more comprehensive range of hydrometeorological and geographical variability. Although this dataset contains catchments on specific regions and thus does not cover the full spectrum of hydrological regimes, it is arguably the best comprehensive catchment scale dataset available currently. These experiments are done similarly to those in \citet{Newman2017}and \citet{Kratzert2021}, keeping the training configuration the same. However, the static catchment attributes and continuous hydrometeorological forcings differ from CAMELS since they are derived from ERA5 and HydroATLAS. The hyperparameters used in LSTMs are taken from \citet{Kratzert2019}, the combination of the best-reported performance. The hyperparameters of TFTs are the same as the best-performing combination found for the CAMELS benchmarking experiment (Table A1).

We train ten LSTMs and ten TFTs with different random initializations. Once the models are trained, we take the mean of discharges (average ensemble) from all ten models and compare the performance of TFTs with LSTMs. Then, we evaluate the variability in the performance of both models with respect to catchment attributes (mean latitude, mean elevation, size, and slope). This analysis is carried out for both CAMELS and Caravan experiments. A key edge of attention-based neural networks, compared to recurrent neural networks, is its ability to learn from longer sequences. They can access the entire sequence at a time and learn from both distant and near past time steps. To investigate this property, we train both LSTMs and TFTs on CAMELS data with different sequence lengths (730, 1095, 1460, and 1825 days). In the benchmarking experiments, we provide sequences of a complete year. If the attention module in TFT can leverage information from distant data, the performance should be better than that of LSTMs as we increase the sequence length. Lastly, we demonstrate the explainability of TFT models with variable importance and relevant past-time steps identified by the model. 

\begin{table}[ht] \centering
\begin{threeparttable}
    \caption{Metrics used to evaluate the performance of models.}
    \begin{tabular}{l >{\raggedright\arraybackslash}p{0.2\linewidth}l >{\raggedright\arraybackslash}p{0.15\linewidth}}
        \toprule
        \textbf{Metric} & \textbf{Description} & \textbf{Equation} & \textbf{Range} \\
        \midrule
        NSE & Nash-Sutcliffe efficiency \citep{Nash1970} & 
        $
        NSE = 1 - \frac{\sum_{i=1}^{n} (Q_{\text{sim},i} - Q_{\text{obs},i})^2}{\sum_{i=1}^{n} (Q_{\text{obs},i} - \overline{Q_{\text{obs}}})^2}
        $
        & $(-\infty, 1]$; closer to 1 is better \\

        KGE & Kling-Gupta efficiency \citep{Gupta2009}& 
        $
        KGE = 1 - \sqrt{(r - 1)^2 + (\alpha - 1)^2 + (\beta - 1)^2}
        $
        & $(-\infty, 1]$; closer to 1 is better \\

        Pearson-$r$ & Pearson correlation coefficient & 
        $
        r = \frac{\sum_{i=1}^{n} (Q_{\text{sim},i} - \overline{Q_{\text{sim}}})(Q_{\text{obs},i} - \overline{Q_{\text{obs}}})}
        {\sqrt{\sum_{i=1}^{n} (Q_{\text{sim},i} - \overline{Q_{\text{sim}}})^2} 
        \sqrt{\sum_{i=1}^{n} (Q_{\text{obs},i} - \overline{Q_{\text{obs}}})^2}}
        $
        & $[-1,1]$; closer to 1 is better \\

        $\alpha$-NSE & Ratio of observed and simulated standard deviations \citep{Gupta2009} & 
        $
        \alpha = \frac{\sigma_{\text{sim}}}{\sigma_{\text{obs}}}
        $
        & $(0, \infty]$; closer to 1 is better \\

        $\beta$-NSE & Ratio of observed and simulated means \citep{Gupta2009} & 
        $\beta = \frac{\mu_{\text{sim}} - \mu_{\text{obs}}}{\mu_{\text{obs}}}$
        & $(-\infty, \infty)$; closer to 0 is better \\

        FHV & Top 2\% peak flow bias \citep{Yilmaz2008} & See Eq. (A3) in \citet{Yilmaz2008} & $(-\infty, \infty)$; closer to 0 is better \\

        FLV & Bottom 30\% low flow bias \citep{Yilmaz2008} & See Eq. (A4) in \citet{Yilmaz2008} & $(-\infty, \infty)$; closer to 0 is better \\

        FMS & Flow duration curve midsegment slope bias \citep{Yilmaz2008} & See Eq. (A2) in \citet{Yilmaz2008} & $(-\infty, \infty)$; closer to 0 is better \\

        Peak timing & Mean peak time lag (days) between observed and simulated peaks \citep{Yilmaz2008} & See Appendix B in \citet{Yilmaz2008} & $(-\infty, \infty)$; closer to 0 is better \\
        \bottomrule
    \end{tabular}
    \label{tab:metrics_def}
\end{threeparttable}
\end{table}

\section{Results and Discussion}
\subsection{Performance on CAMELS}
After training ten randomly initialized TFT and LSTM models on the data of 531 CAMELS catchment, we obtained hydrographs from averaging the outputs of ten models (average ensemble). The evaluation metrics estimated from these results, as shown in Table 2, suggest that the performance of TFT is slightly better than LSTM, although not significantly different. While the median test KGE of TFT is 0.801, LSTM has 0.796. Similarly, the median test NSE of TFT is 0.821, and of LSTM is 0.820. However, there is an improvement in the top 2\% peak flow bias (FHV) and flow duration curve midsegment slope bias (FMS) in the case of TFT. This suggests that TFT is able to capture peaks and midsegments of the hydrograph better than LSTM. In addition, compared to LSTM, the number of catchments showing greater than 0.8 KGE is increased by 13 with TFT (Figure 2a). The additional attention layer in the TFT could be the potential reason for these improvements. In addition to the local context learning ability of recurrent layer, TFT can simultaneously attend to all the peaks, rising limbs, and falling limbs to learn the rainfall to runoff process that drives the processes driving peaks and midsegment of hydrographs.

We examined the variability of KGE with respect to four catchment attributes: 1) mean latitude, 2) size or area, 3) mean elevation, and 4) mean slope of the catchment. The performance of both TFT and LSTM improves with respect to the mean latitude and elevation (although not evidently as of latitude) of the catchment (Figure 3 a and c). The potential reason for this outcome could be that the influence of snow processes and soil moisture at higher latitudes and elevations increases the predictability of runoff \citep{Lettenmaier1994, Maurer2004}. Interestingly, in terms of KGE distribution, the outperformance of TFT over LSTM increases with respect to the mean elevation and size of the catchments. This is most likely because of the hydrological response time of larger catchments, where water must travel longer distances, and high-altitude catchments, where gradual snow processes majorly drive the runoff generation. In such cases, TFT can leverage its attention mechanism to learn from prolonged processes, as it can attend to any part of the sequence (see section 3.3. Long-term learning for more discussion).

In addition to average ensembling, for comparison, we also obtained hydrographs from the best-performing model (based on validation data) out of ten randomly initialized models (Figure 2b). As shown in Table 2, almost all evaluation metrics are dropping compared to the average ensemble. This could be because of the overdependence on one single model for prediction in the latter case. When the predictions of multiple models are averaged, the errors made by models tend to be canceled by each other. This reduces the variance in the overall prediction without substantial change in bias, resulting in a better bias-variance tradeoff. These factors do not come into play when predictions are based on a single model, regardless of its best performance on validation data. 

\begin{table}[ht]
\centering
\begin{threeparttable}
    \caption{Performance of TFT and LSTM trained with CAMELS data. Average ensemble denotes the metrics obtained from averaging the predictions of ten randomly initialized models. Best of ten denotes that obtained from the predictions of one model, out of ten that performed best on validation data. }
    
    \begin{tabular}{l c c c c}
        \toprule
        Metric & \multicolumn{2}{c}{Average ensemble} & \multicolumn{2}{c}{Best of ten} \\
        \cmidrule(lr){2-3} \cmidrule(lr){4-5}
        & LSTM & TFT & LSTM & TFT \\
        \midrule
        NSE         & 0.820  & 0.821  & 0.792  & 0.798  \\
        KGE         & 0.796  & 0.801  & 0.789  & 0.785  \\
        Pearson r   & 0.915  & 0.915  & 0.903  & 0.901  \\
        $\alpha$-NSE & 0.853  & 0.867  & 0.860  & 0.850  \\
        $\beta$-NSE & -0.030 & -0.023 & -0.032 & -0.028 \\
        FHV         & -14.602 & -12.906 & -13.913 & -13.679 \\
        FLV         & 34.067  & 35.030  & 1.390   & 32.557 \\
        FMS         & -8.200  & -7.365  & -7.821  & -9.754 \\
        Peak timing & 0.238  & 0.238  & 0.267  & 0.261  \\
        \bottomrule
    \end{tabular}
    \label{tab:metrics_vals}
\end{threeparttable}
\end{table}

\begin{figure}
	\centering
    \includegraphics[width=0.7\textwidth]{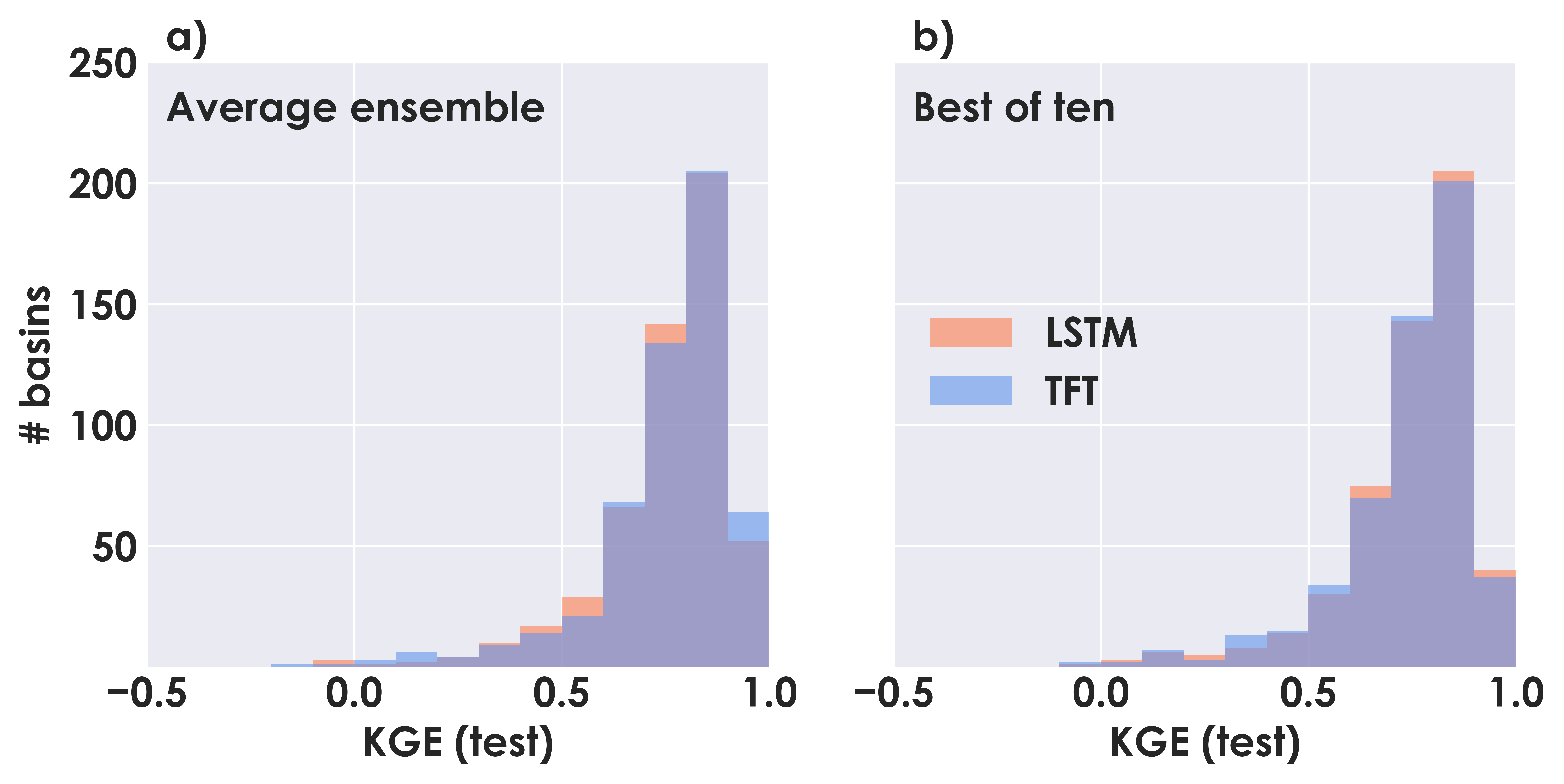}
	\caption{Distribution of TFT and LSTM KGEs on test split of CAMELS data. a) KGEs obtained from averaging the predictions of ten randomly initialized models. b) KGEs obtained from the predictions of one model, out of ten that performed best on validation data.}
	\label{fig:fig2}
\end{figure}

\begin{figure}
	\centering
    \includegraphics[width=0.7\textwidth]{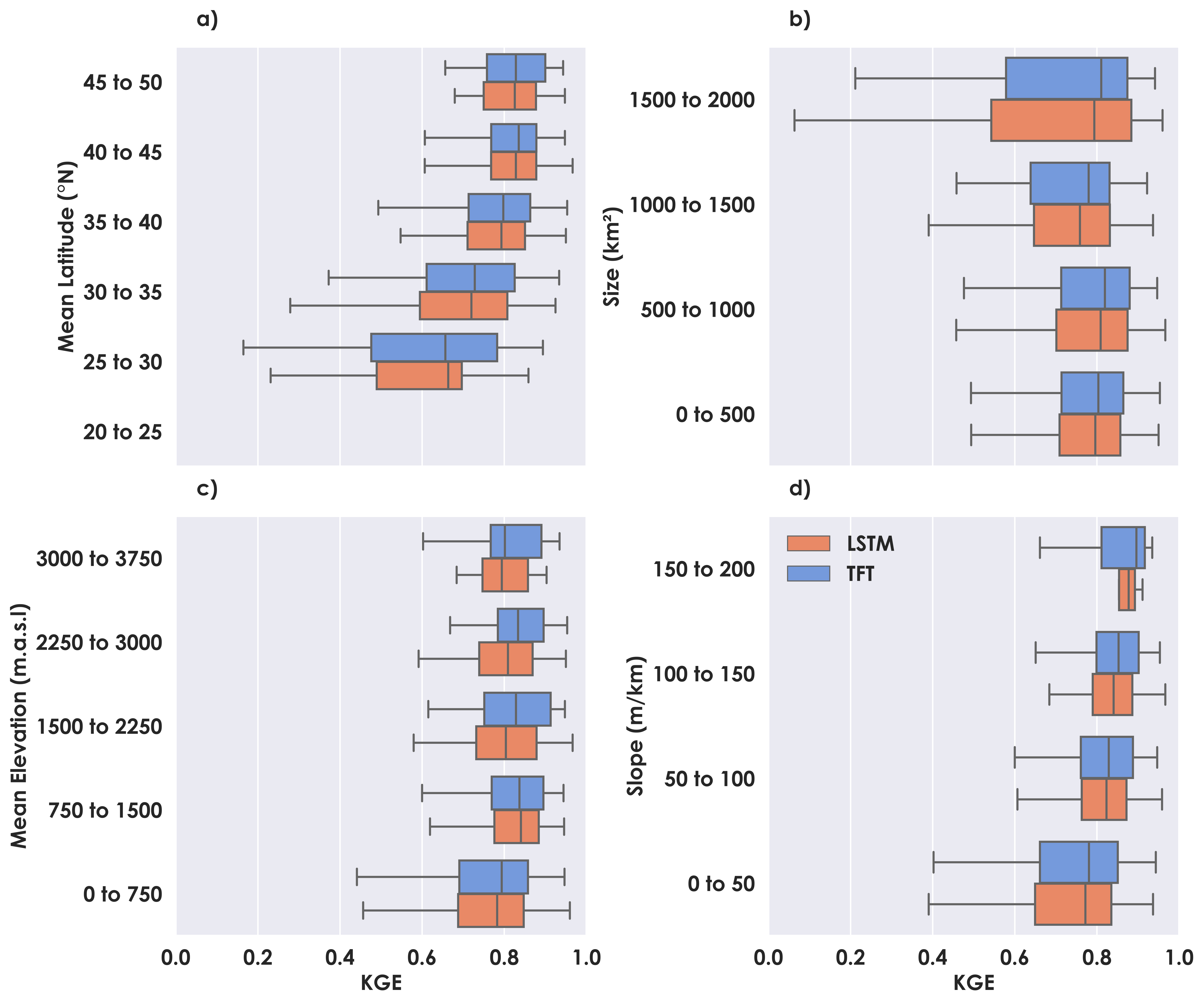}
	\caption{Variability of model performance with respect to a) mean latitudes, b) size, c) mean elevation, and d) slope of the catchments in CAMELS.}
	\label{fig:fig3}
\end{figure}

\subsection{Performance on Caravan}
We repeated the same experiments on the Caravan data in five subsets (United States; US, Brazil; BR, Great Britain; GB, Australia; AUS, Chile; CL). The results (Table 3) suggest that, compared to the original CAMELS, nearly all metrics drop when trained with forcings from Caravan, regardless of the higher number of input variables. This can be due to the significant bias in forcing variables in ERA5 land, a global reanalysis product, where the regional process might not be represented accurately \citep{Beck2017}. Consequently, this could deteriorate the performance of hydrological modeling, as reported by several previous studies \citep{MariaClerc-Schwarzenbach2024, Tarek2020}. In general, our outcome suggests that the Caravan dataset comes with caveats to use for rainfall-runoff modeling.

When we evaluate the performance of TFT and LSTM, we see that the NSEs of TFT are slightly better than those of LSTM, although not significantly different, in the case of the US, Brazil, and Great Britain. At the same time, there is a slight decrease in KGE for these three regions. LSTM has marginally better scores than TFT for Australia, whereas Chile has no difference. Given the questionable fitness of forcings in the Caravan dataset, it is difficult to argue that one model is advantageous over another.

The variability of model performance with respect to mean latitude, size, and mean elevation of catchments is shown in Figure 4 a, b, and c, respectively. As we go towards poles, both TFT and LSTM improve performance. As pointed out in the last section, this might be due to the larger influence of snow processes and soil moisture and the resulting enhanced predictability of runoff \citep{Lettenmaier1994, Maurer2004}. For the same reasons, both models improve the performance as elevation increases, as shown in Figure R4c. Note that there is a sudden drop in KGE for the catchments higher than 4000 from the mean sea level. Although the hydrometeorological processes in these elevations can be complex, this further raises the issue of the inadequacy of the dataset in representing these processes. Nevertheless, we can observe that the outperformance of TFT over LSTM increases as elevation increases, as we observed in the CAMELS-based experiment. This can further reassure that the TFT can leverage its attention mechanism to learn from slower hydrological processes (e.g., snow processes). 

\begin{table}[h] \centering
\begin{threeparttable}
    \caption{Performance of TFT and LSTM trained with Caravan data in subsets (United States; US, Brazil; BR, Great Britain; GB, Australia; AUS, Chile; CL). The metrics are obtained from the average of predictions of ten randomly initialized models.}
    
    \begin{tabular}{l cc cc cc cc cc}
        \toprule
        \textbf{Metric} & \multicolumn{2}{c}{\textbf{CAMELS-US}} & \multicolumn{2}{c}{\textbf{CAMELS-BR}} & \multicolumn{2}{c}{\textbf{CAMELS-GB}} & \multicolumn{2}{c}{\textbf{CAMELS-AUS}} & \multicolumn{2}{c}{\textbf{CAMELS-CL}} \\
        & LSTM & TFT & LSTM & TFT & LSTM & TFT & LSTM & TFT & LSTM & TFT \\
        \midrule
        NSE  & 0.54 & 0.56 & 0.43 & 0.45 & 0.77 & 0.78 & 0.58 & 0.57 & 0.75 & 0.75 \\
        KGE  & 0.67 & 0.66 & 0.59 & 0.56 & 0.83 & 0.82 & 0.61 & 0.55 & 0.75 & 0.75 \\
        Pearson-$r$ & 0.77 & 0.77 & 0.70 & 0.70 & 0.89 & 0.89 & 0.79 & 0.78 & 0.89 & 0.89 \\
        $\alpha$-NSE & 0.89 & 0.86 & 0.80 & 0.72 & 0.93 & 0.90 & 0.77 & 0.67 & 0.91 & 0.91 \\
        $\beta$-NSE & 0.01 & 0.01 & 0.05 & 0.01 & 0.05 & 0.02 & -0.00 & -0.02 & -0.00 & -0.00 \\
        FHV  & -10.77 & -14.11 & -21.05 & -28.72 & -7.73 & -12.43 & -22.10 & -28.96 & -6.11 & -7.74 \\
        FLV  & 36.21 & 24.81 & 38.50 & 33.18 & 23.81 & 14.21 & -4.22 & 15.44 & 52.70 & 55.97 \\
        FMS  & -14.14 & -13.03 & -9.50 & -9.04 & -6.88 & -6.61 & -16.76 & -18.47 & -11.13 & -8.15 \\
        Peak timing & 0.40 & 0.39 & 1.00 & 1.00 & 0.30 & 0.29 & 0.42 & 0.46 & 0.51 & 0.50 \\
        \bottomrule
    \end{tabular}
    \label{tab:performance}
\end{threeparttable}
\end{table}

\begin{figure}
	\centering
    \includegraphics[width=0.7\textwidth]{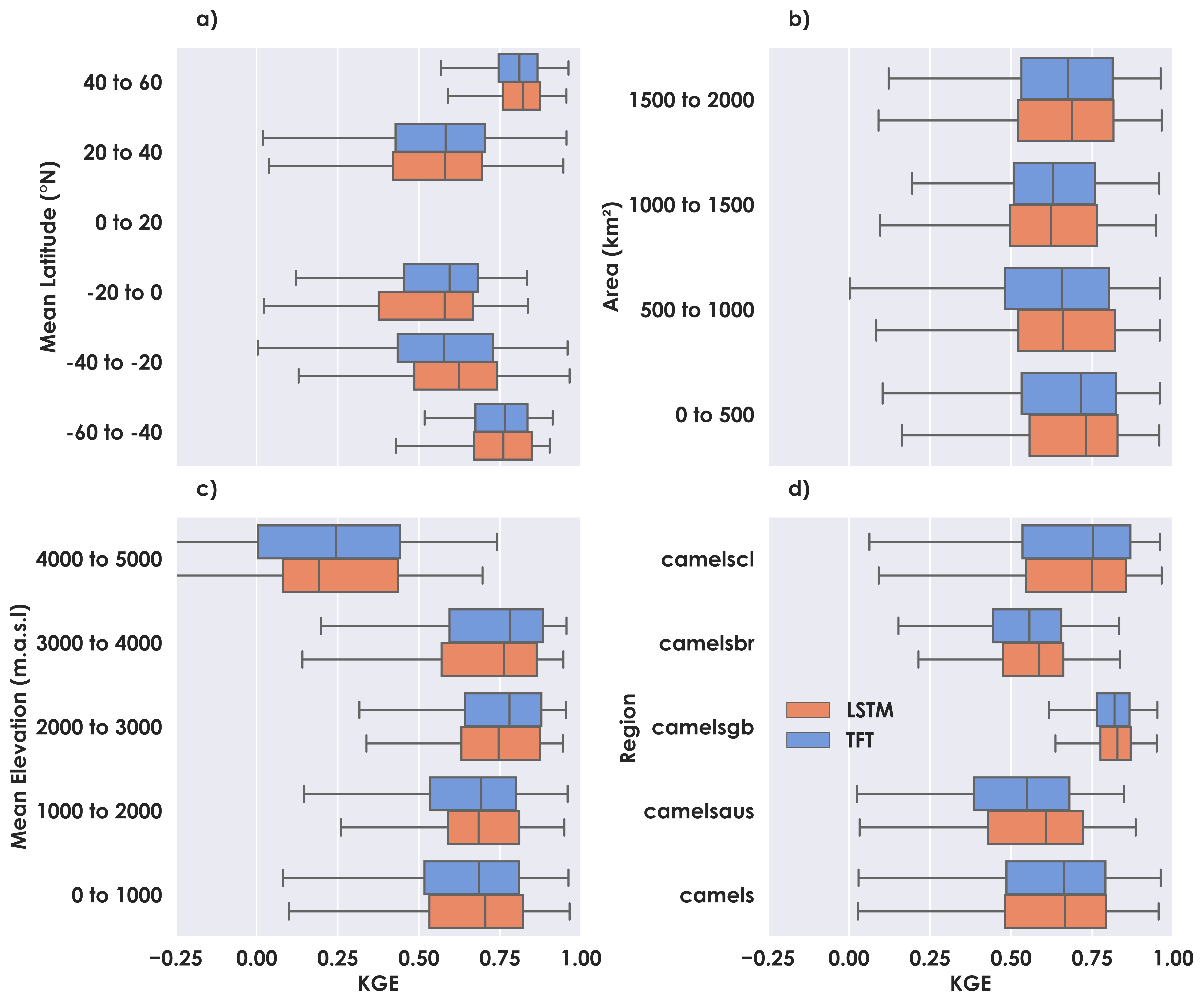}
	\caption{Variability of model performance with respect to a) mean latitudes, b) area or size, c) mean elevation of the catchments in Caravan. d) Performance distribution in five Caravan subsets.}
	\label{fig:fig4}
\end{figure}

\subsection{Long-term learning}
We trained both LSTM and TFT with different sequence lengths to see if the model can learn the rainfall-runoff process better with longer sequences. As shown in Figure 5, the distribution of KGEs from LSTM remains nearly identical as we increase the sequence length. However, for TFT, the distribution improves as higher sequence lengths are used. This validates our finding that, using its attention mechanism, TFT can learn from distant data better than LSTM. It can be noted that the performance went down at a sequence length of 1,825 days (five years). A potential reason for this decline is the decreased number of training sequences. At a sequence length of five years, the total number of sequences is reduced nearly to half (note that the total training period is ten years), compared to a sequence of one year.

A direct proof of this ability can be observed in the distribution of the attention scores that TFT computed for past timesteps in the sequence (Figure 6). We can see that the distribution gets wider in the preceding months, and the pattern repeats for the same months in previous years. Due to seasonality in hydrometeorological processes, the runoff-generating processes can have similar patterns on the same days as in previous years. Therefore, the model can take advantage of these patterns to better learn the rainfall-runoff process. This is what the distribution of attention scores by TFT reveals. We argue that, given the step-by-step processing of data, LSTM might not be able to benefit from the information in the patterns existing in long sequences. 

\begin{figure}
	\centering
    \includegraphics[width=0.7\textwidth]{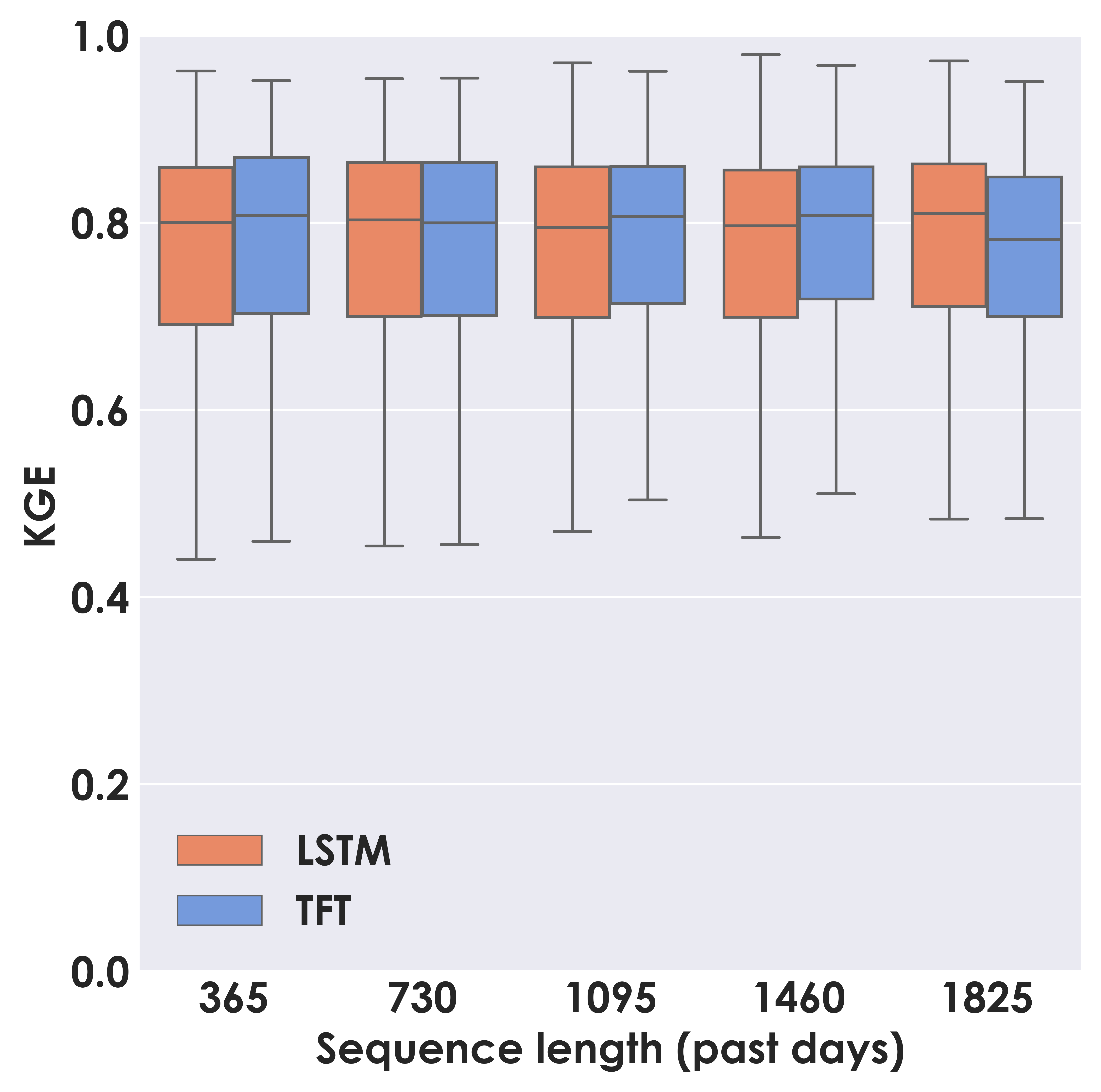}
	\caption{Performance distribution of LSTM and TFT for different training sequence lengths.}
	\label{fig:fig5}
\end{figure} 

\begin{figure}
	\centering
    \includegraphics[width=0.7\textwidth]{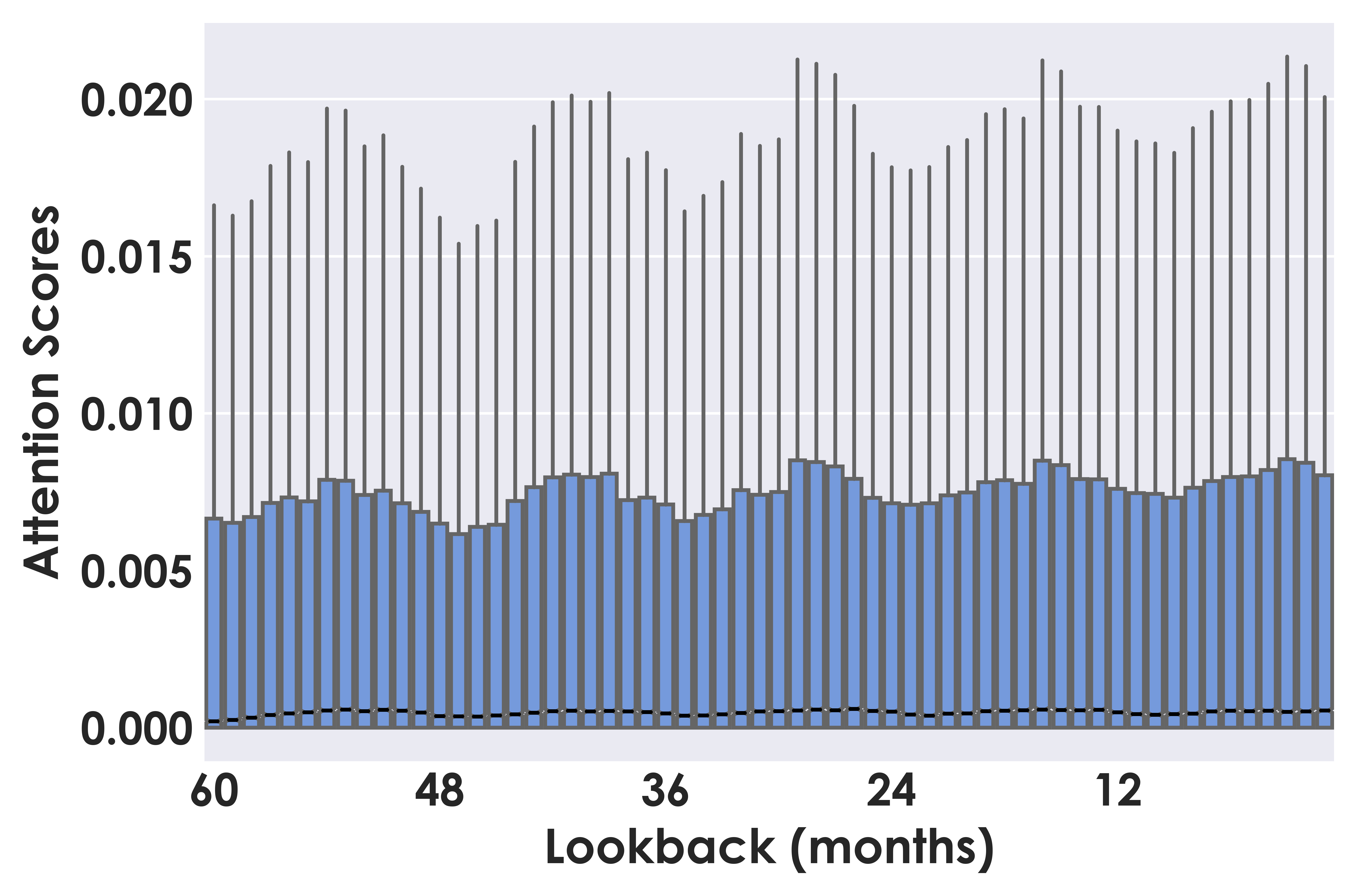}
	\caption{Distribution of relative attention scores. Each boxplot represents the distribution of attention scores used by TFT on input test sequences for months prior to the predicting time step.}
	\label{fig:fig6}
\end{figure}

\subsection{Interpretability of TFT}
Possibly, the key advantage of TFT is its explainability. The model architecture is built in such a way that we can gain some insights into why it predicted the predicted values. Two components in the architecture can be used for this purpose: 1) the interpretable multi-head attention block and 2) the variable selection networks. We have already discussed the former in the previous section (Figure 6), where we revealed the previous time steps the model emphasizes (or attends) in learning the rainfall-runoff process. In this section, we will focus on the variable selection networks.

TFT assigns weights to each input variable, both static and dynamic, before passing on to intermediate blocks. We can extract these weights for each sample to gain information about the relative importance of input variables for prediction. Figure 7a shows the distribution of weights assigned by TFT to each dynamic input variable used for predicting the test data in the CAMELS experiment. We can see that precipitation is more important than other variables. This is reasonably expected as precipitation is the only incoming water flux to catchments, and it also validates that the TFT can learn information agreeing to the existing knowledge about the rainfall-runoff process. Interestingly, we can also observe that out of the precipitation data source, Daymet is generally given more importance, followed by Maurer and NLDAS. In other words, TFT relies on Daymet precipitation more than others to predict the discharge. This indicates that Daymet might have reliable and representative precipitation information from a rainfall-runoff modeling perspective, whereas NLDAS could be the opposite. Such findings potentially show the relative quality of forcing sources.

After precipitation, the most weighted variable is the shortwave radiation. This is most likely because of the role of shortwave radiation in the snow processes. Shortwave radiation is the incoming energy flux to the earth system, which can be crucial in driving the snow processes \citep{RasiyaKoya2023}. An increased intensity of solar radiation can exacerbate snow melting. Here, it should be noted that the minimum and maximum temperatures do not carry much importance compared to shortwave radiation. This is likely because the information about energy-driven processes can be better captured from shortwave radiation as it is a more direct measure of energy rather than temperature, which is an outcome of the shortwave radiation. We can also observe the relative importance of forcing sources from the weights TFT gives. In the case of shortwave radiation, the model depends more on data from NLDAS to produce discharge, which indicates that NLDAS might have a better representation of shortwave radiation. Such analysis can also be extended to other variables to assess the reliability of forcing data sources. 
Besides dynamic data, we can find the relative importance of static attributes of the catchment as well from TFT (Figure 7b). As per weights assigned by TFT, the most important static variable in predicting the discharge of a catchment is maximum water content (i.e., the maximum amount of water that can be contained in a unit mass or volume of soil). This makes sense as the maximum water content of the soil is a crucial threshold for determining the incoming rainfall to produce direct runoff or infiltration, a bifurcation significantly contributing to the discharge \citep{Liu2011}. Mean precipitation and leaf area index difference are the static variables that come next in terms of relative variable importance. The long-term mean precipitation can provide information about the stable incoming flux of water, which can aid the model in determining the baseflow in the hydrograph. Leaf area index difference, a measure of the vegetation variability in the catchment, can potentially assist the model in learning information about the canopy intercepted water and magnitudes of evapotranspiration \citep{Leuning2008}. Both these variables are vital in the water balance of catchments.

\begin{figure}
	\centering
    \includegraphics[width=0.9\textwidth]{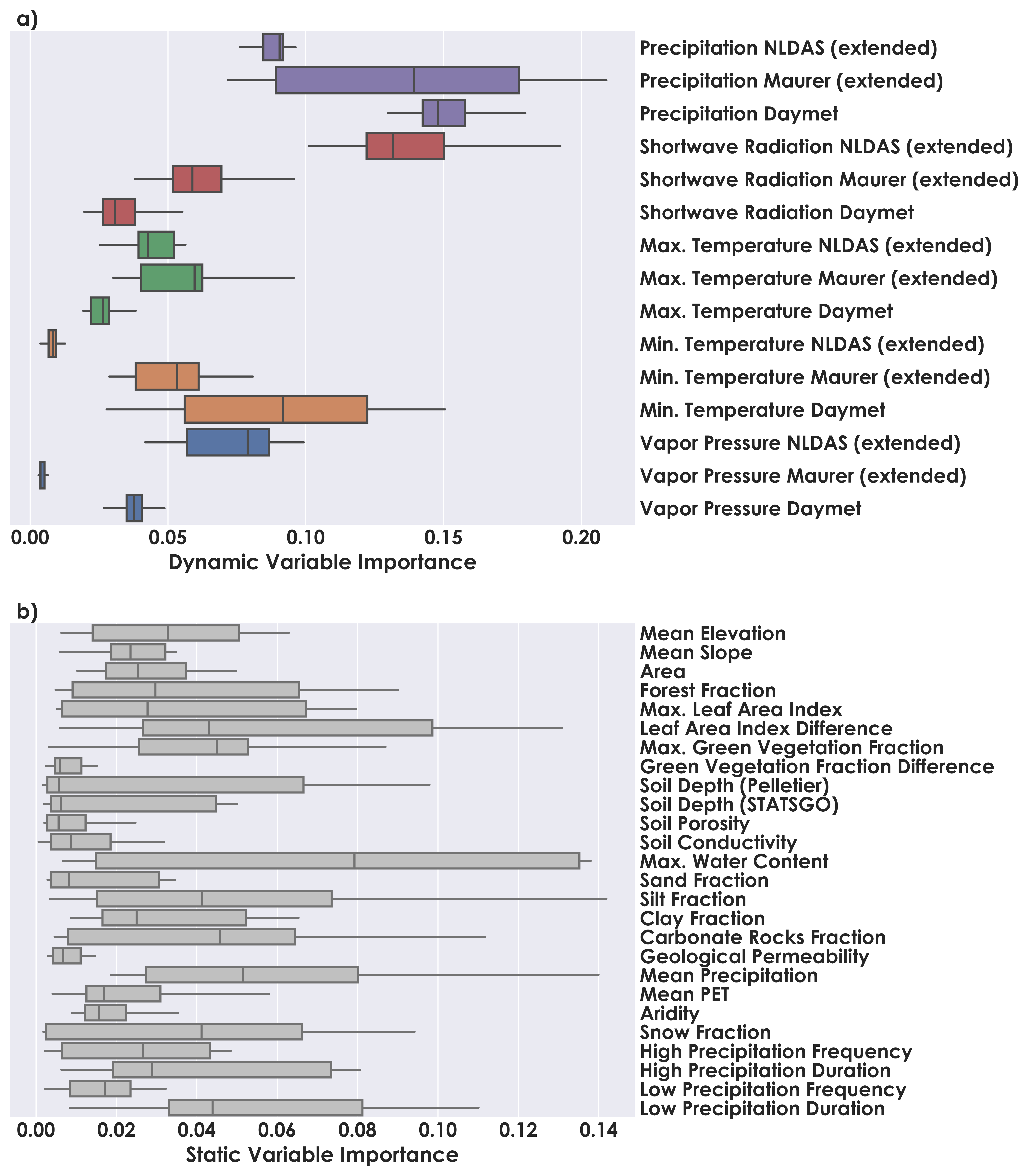}
	\caption{Variable importance plot from TFT}
	\label{fig:fig7}
\end{figure}

\section{Conclusion}
We train single TFT and LSTM models for rainfall-runoff modeling in 531 CAMELS catchments, repeat the setup with ten random initializations, and take the average ensemble. Then, the experiment is repeated with five subsets of the Caravan dataset. We analyze and compare the performance of TFT and LSTM on both CAMELS and Caravan datasets. The variability of performance regarding the catchment attributes is also examined. Both models are then trained on CAMELS with different sequence lengths to show the effect of long sequences on model performance. Finally, we demonstrate how we can gain scientific insights from the explainability of the TFT architecture.

Our results show that TFT performs slightly better than LSTM for rainfall-runoff simulation with the CAMELS dataset. Majorly, TFT improves the simulation of the peak and midsection of the hydrographs. TFT can be the better candidate for catchments at elevations, latitudes, or with large sizes or slopes. TFT can exploit longer sequences better by attending to seasonally repeating hydroclimatic conditions of the past years. TFT reveals precipitation and shortwave radiation as important dynamic input variables. Similarly, the maximum water content of the soil is identified as the most important static variable. Moreover, the model identifies Daymet for precipitation and NLDAS for shortwave radiation as reliable sources to simulate discharge at the outlet. Interestingly, for the same 531 catchments, the performance of TFT and LSTM significantly reduces when trained with the Caravan dataset, pointing out its potential data quality issues. 

\section*{Acknowledgments}
\addcontentsline{toc}{section}{Acknowledgments}
We thank Python, PyTorch \citep{Paszke2019}, and Neuralhydrology \citep{kratzert2022joss} for providing the tools to set up the models and run our experiments. We are grateful to Holland Computing Center at the University of Nebraska-Lincoln for providing the computing resources. We extend our thanks to Dr. Grey Nearing for the valuable discussion and feedback about our efforts in model setup. 
\newpage

\appendix
\section{Hyperparameter tuning}
\renewcommand{\thetable}{\Alph{section}\arabic{table}} 
\setcounter{table}{0}

\begin{table}[H] \centering
\begin{threeparttable}
    \caption{Hyperparameter tuning of TFT on CAMELS benchmarking experiment. Bolded values correspond to the best-performing combination.}
    \begin{tabular}{l p{9cm} c}
        \toprule
        \textbf{Hyperparameter} & \textbf{Description} & \textbf{Values} \\
        \midrule
        Attention head size & Number of attention heads in the multi-head attention block. & \textbf{4}, 8, 16 \\
        Dropout & Fraction of deactivating random neurons in GRN, LSTM, and multi-head attention blocks. & \textbf{0.1}, 0.2, 0.3 \\
        Hidden size & Hidden size for variable selection, LSTM, GRN, and attention blocks. & 16, \textbf{64}, 256 \\
        LSTM layers & Number of LSTM layers. & 1, \textbf{2} \\
        Batch size & Number of training examples in one iteration. & 64, 128, \textbf{264} \\
        \bottomrule
    \end{tabular}
    \label{tab:hyperparams}
\end{threeparttable}
\end{table}

\newpage

\section{CAMELS variables}
\setcounter{table}{0}

\begin{table}[H] \centering
\begin{threeparttable}
    \caption{List of the CAMELS variables. Adapted from \citet{Addor2017}, \citet{Kratzert2019}, and \citet{Newman2015}.}
    \begin{tabular}{c l l}
        \toprule
        \textbf{Category} & \textbf{Variable Name} & \textbf{Description} \\
        \midrule
        \multirow{15}{*}{\rotatebox{90}{\parbox{5cm}{\centering Hydrometeorological forcings \\ (continuous daily time series)}}} & PRCP(mm/day)\_nldas\_extended & Precipitation from extended NLDAS forcing \\
        & SRAD(W/m$^2$)\_nldas\_extended & Solar radiation from extended NLDAS forcing \\
        & Tmax(C)\_nldas\_extended & Maximum temperature from extended NLDAS forcing \\
        & Tmin(C)\_nldas\_extended & Minimum temperature from extended NLDAS forcing \\
        & Vp(Pa)\_nldas\_extended & Vapor pressure from extended NLDAS forcing \\
        & prcp(mm/day)\_maurer\_extended & Precipitation from extended Maurer forcing \\
        & srad(W/m$^2$)\_maurer\_extended & Solar radiation from extended Maurer forcing \\
        & tmax(C)\_maurer\_extended & Maximum temperature from extended Maurer forcing \\
        & tmin(C)\_maurer\_extended & Minimum temperature from extended Maurer forcing \\
        & vp(Pa)\_maurer\_extended & Vapor pressure from extended Maurer forcing \\
        & prcp(mm/day)\_daymet & Precipitation from NLDAS forcing \\
        & srad(W/m$^2$)\_daymet & Solar radiation from NLDAS forcing \\
        & tmax(C)\_daymet & Maximum temperature from NLDAS forcing \\
        & tmin(C)\_daymet & Minimum temperature from NLDAS forcing \\
        & vp(Pa)\_daymet & Vapor pressure from NLDAS forcing \\
        \midrule
        \multirow{25}{*}{\rotatebox{90}{\parbox{5cm}{\centering Catchment attributes \\ (static covariates)}}} & elev\_mean & Catchment mean elevation. \\
        & slope\_mean & Catchment mean slope. \\
        & area\_gages2 & Catchment area. \\
        & frac\_forest & Forest fraction. \\
        & lai\_max & Maximum monthly mean of leaf area index. \\
        & lai\_diff & Difference between max. and min. monthly mean of leaf area index. \\
        & gvf\_max & Maximum monthly mean of green vegetation fraction. \\
        & gvf\_diff & Difference between max. and min. monthly mean of green vegetation fraction. \\
        & soil\_depth\_pelletier & Depth to bedrock (maximum 50m). \\
        & soil\_depth\_statsgo & Soil depth (maximum 1.5m). \\
        & soil\_porosity & Volumetric porosity. \\
        & soil\_conductivity & Saturated hydraulic conductivity. \\
        & max\_water\_content & Maximum water content of soil. \\
        & sand\_frac & Fraction of sand in soil. \\
        & silt\_frac & Fraction of silt in soil. \\
        & clay\_frac & Fraction of clay in soil. \\
        & carbonate\_rocks\_frac & Fraction of catchment area characterized as carbonate sedimentary rocks. \\
        & geol\_permeability & Surface permeability (log10). \\
        & p\_mean & Mean daily precipitation. \\
        & pet\_mean & Mean daily potential evapotranspiration. \\
        & aridity & Ratio of mean PET to mean precipitation. \\
        & frac\_snow & Fraction of precipitation on days with temperature below 0°C. \\
        & high\_prec\_freq & Frequency of high-precipitation days ($\geq$ 5 times mean daily precipitation). \\
        & high\_prec\_dur & Average duration of high-precipitation events (days). \\
        & low\_prec\_freq & Frequency of dry days (<1 mm/day). \\
        & low\_prec\_dur & Average duration of dry periods (days with precipitation <1 mm). \\
        \bottomrule
    \end{tabular}
    \label{tab:hydro_catchment}
\end{threeparttable}
\end{table}

\newpage
\setcounter{table}{0}

\section{Caravan variables}
\begin{table}[h!] \centering
\begin{threeparttable}
    \caption{List of dynamic variables. Note that since minimum net solar radiation is a constant (equal to zero), it is considered a static variable (this difference only affects the case of TFT due to the model architecture where static covariates are processed separately). Table adapted from \citet{Kratzert2023a}and \citet{RasiyaKoya2023a}}
    
    \begin{tabular}{lll}
    \toprule
    \textbf{Variable Name} & \textbf{Description} & \textbf{Aggregation} \\
    \midrule
    snow\_depth\_water\_equivalent\_mean & Snow-Water Equivalent in mm & Daily min, max, mean \\
    surface\_net\_solar\_radiation\_mean & Surface net solar radiation in W/m$^2$ & Daily min, max, mean \\
    surface\_net\_thermal\_radiation\_mean & Surface net thermal radiation in W/m$^2$ & Daily min, max, mean \\
    surface\_pressure\_mean & Surface pressure in kPa & Daily min, max, mean \\
    temperature\_2m\_mean & 2m air temperature in °C & Daily min, max, mean \\
    dewpoint\_temperature\_2m\_mean & 2m dew point temperature in °C & Daily min, max, mean \\
    u\_component\_of\_wind\_10m\_mean & U-component of wind at 10m in m/s & Daily min, max, mean \\
    v\_component\_of\_wind\_10m\_mean & V-component of wind at 10m in m/s & Daily min, max, mean \\
    volumetric\_soil\_water\_layer\_1\_mean & Volumetric soil water layer 1 (0-7cm) in m$^3$/m$^3$ & Daily min, max, mean \\
    volumetric\_soil\_water\_layer\_2\_mean & Volumetric soil water layer 2 (7-28cm) in m$^3$/m$^3$ & Daily min, max, mean \\
    volumetric\_soil\_water\_layer\_3\_mean & Volumetric soil water layer 3 (28-100cm) in m$^3$/m$^3$ & Daily min, max, mean \\
    volumetric\_soil\_water\_layer\_4\_mean & Volumetric soil water layer 4 (100-289cm) in m$^3$/m$^3$ & Daily min, max, mean \\
    total\_precipitation\_sum & Total precipitation in mm & Daily sum \\
    potential\_evaporation\_sum & Potential evapotranspiration in mm & Daily sum \\
    streamflow & Observed streamflow in mm/d & Daily mean \\
    \bottomrule
    \end{tabular}
    \label{tab:dynvars} 
\end{threeparttable}
\end{table}

\begin{table}[ht!]\centering
\begin{threeparttable}
    \caption{List of static covariates. Refer to \citet{Kratzert2023a} and the "BasinATLAS Catalogue" of HydroATLAS (https://www.hydrosheds.org/hydroatlas) for the explanation of features. Table adapted from \citet{RasiyaKoya2023a}.}

    \begin{tabular}{lllllll}
        \toprule
        \multicolumn{7}{c}{Static Covariate Name} \\
        \midrule
        p\_mean & glc\_pc\_s13 & cmi\_ix\_syr & inu\_pc\_smn & pet\_mean & prm\_pc\_sse & pet\_mm\_s11 \\
        pnv\_cl\_smj & aridity & glc\_pc\_s14 & pet\_mm\_s12 & pre\_mm\_s08 & frac\_snow & glc\_pc\_s11 \\
        pet\_mm\_s10 & pre\_mm\_s09 & moisture\_index & glc\_pc\_s12 & tmp\_dc\_smn & run\_mm\_syr & seasonality \\
        glc\_pc\_s10 & wet\_pc\_s08 & pre\_mm\_s06 & high\_prec\_freq & kar\_pc\_sse & wet\_pc\_s09 & pre\_mm\_s07 \\
        high\_prec\_dur & slp\_dg\_sav & slt\_pc\_sav & pre\_mm\_s04 & low\_prec\_freq & glc\_pc\_s19 & wet\_pc\_s02 \\
        pre\_mm\_s05 & low\_prec\_dur & tmp\_dc\_s07 & wet\_pc\_s03 & snd\_pc\_sav & sgr\_dk\_sav & tmp\_dc\_s08 \\
        wet\_pc\_s01 & pre\_mm\_s02 & glc\_pc\_s06 & tmp\_dc\_s05 & hdi\_ix\_sav & pre\_mm\_s03 & glc\_pc\_s07 \\
        tmp\_dc\_s06 & wet\_pc\_s06 & ele\_mt\_sav & nli\_ix\_sav & tmp\_dc\_s09 & wet\_pc\_s07 & pre\_mm\_s01 \\
        glc\_pc\_s04 & for\_pc\_sse & wet\_pc\_s04 & urb\_pc\_sse & glc\_pc\_s05 & aet\_mm\_s06 & wet\_pc\_s05 \\
        lka\_pc\_sse & glc\_pc\_s02 & aet\_mm\_s05 & fec\_cl\_smj & pre\_mm\_s10 & rev\_mc\_usu & aet\_mm\_s08 \\
        glc\_cl\_smj & dis\_m3\_pmx & glc\_pc\_s03 & aet\_mm\_s07 & swc\_pc\_syr & snw\_pc\_s01 & glc\_pc\_s01 \\
        aet\_mm\_s09 & hft\_ix\_s09 & snw\_pc\_s02 & pet\_mm\_syr & tmp\_dc\_s10 & soc\_th\_sav & snw\_pc\_s03 \\
        dor\_pc\_pva & tmp\_dc\_s11 & gdp\_ud\_sav & snw\_pc\_s04 & glc\_pc\_s08 & aet\_mm\_s02 & dis\_m3\_pyr \\
        snw\_pc\_s05 & glc\_pc\_s09 & aet\_mm\_s01 & gdp\_ud\_ssu & snw\_pc\_s06 & swc\_pc\_s09 & tmp\_dc\_s12 \\
        tmp\_dc\_smx & gla\_pc\_sse & ele\_mt\_smx & aet\_mm\_s04 & cly\_pc\_sav & snw\_pc\_s07 & tbi\_cl\_smj \\
        aet\_mm\_s03 & pet\_mm\_s02 & snw\_pc\_s08 & swc\_pc\_s01 & lit\_cl\_smj & pet\_mm\_s03 & snw\_pc\_s09 \\
        swc\_pc\_s02 & tmp\_dc\_s03 & pet\_mm\_s01 & dis\_m3\_pmn & swc\_pc\_s03 & tmp\_dc\_s04 & riv\_tc\_usu \\
        inu\_pc\_smx & pre\_mm\_s11 & swc\_pc\_s04 & tmp\_dc\_s01 & snw\_pc\_smx & pre\_mm\_s12 & swc\_pc\_s05 \\
        tmp\_dc\_s02 & ppd\_pk\_sav & cmi\_ix\_s07 & swc\_pc\_s06 & cls\_cl\_smj & pet\_mm\_s08 & cmi\_ix\_s08 \\
        swc\_pc\_s07 & pre\_mm\_syr & aet\_mm\_s11 & cmi\_ix\_s05 & swc\_pc\_s08 & pnv\_pc\_s01 & pet\_mm\_s09 \\
        cmi\_ix\_s06 & crp\_pc\_sse & pnv\_pc\_s04 & aet\_mm\_s10 & cmi\_ix\_s09 & glc\_pc\_s22 & pnv\_pc\_s05 \\
        pet\_mm\_s06 & glc\_pc\_s20 & pnv\_pc\_s02 & pet\_mm\_s07 & snw\_pc\_s10 & glc\_pc\_s21 & rdd\_mk\_sav \\
        aet\_mm\_s12 & snw\_pc\_s11 & wet\_pc\_sg1 & ele\_mt\_smn & pet\_mm\_s04 & snw\_pc\_s12 & wet\_pc\_sg2 \\
        pnv\_pc\_s03 & pet\_mm\_s05 & cmi\_ix\_s03 & pac\_pc\_sse & pnv\_pc\_s08 & inu\_pc\_slt & cmi\_ix\_s04 \\
        swc\_pc\_s10 & pnv\_pc\_s09 & ero\_kh\_sav & cmi\_ix\_s01 & swc\_pc\_s11 & pnv\_pc\_s06 & aet\_mm\_syr \\
        cmi\_ix\_s02 & swc\_pc\_s12 & pnv\_pc\_s07 & cmi\_ix\_10 & pst\_pc\_sse & clz\_cl\_smj & wet\_cl\_smj \\
        cmi\_ix\_11 & agg\_area\_frac & gwt\_cm\_sav & snw\_pc\_syr & cmi\_ix\_12 & pop\_ct\_usu & gauge\_lat \\
        gauge\_lon & area & glc\_pc\_s17 & pnv\_pc\_s11 & ari\_ix\_sav & tmp\_dc\_syr & tec\_cl\_smj \\
        ria\_ha\_usu & lkv\_mc\_usu & fmh\_cl\_smj & \\
        \bottomrule
    \end{tabular}
    \label{tab:table}
\end{threeparttable}
\end{table}

\newpage

\bibliographystyle{unsrtnat}


\end{document}